# Detecting the pterygomaxillary disjunction using an instrumented hammer: a cadaveric study

Manon Bas dit Nugues[1], Giuseppe Rosi[2], Charles Henri Flouzat-Lachaniette[3], Gaoussou Toure[4] and Guillaume Haiat[1,*]

[1] CNRS, Univ Paris Est Creteil, Univ Gustave Eiffel, UMR 8208, MSME, 61, Avenue du Général de Gaulle, 94010 Créteil Cedex, France

[2] Univ Paris Est Creteil, Univ Gustave Eiffel, CNRS, UMR 8208, MSME, 61 Avenue du Général de Gaulle, 94010 Créteil Cedex, France

[3] INSERM U955, IMRB Université Paris-Est, 51 avenue du Maréchal de Lattre de Tassigny, 94000 Créteil, France; Service de Chirurgie Orthopédique et Traumatologique, Hôpital Henri Mondor AP-HP, CHU Paris 12, Université Paris-Est, 51 avenue du Maréchal de Lattre de Tassigny, 94000 Créteil, France

[4] Service de chirurgie Maxillofaciale CHI Villeneuve Saint Georges, France

* Correspondance : guillaume.haiat@cnrs.fr ;

Abstract: 233 Words

Main text: 5195 Words

# Abstract

*Background:* In many maxillofacial surgeries, the surgeon must perform a pterygomaxillary disjunction in order to separate the jaw from the skull. The disjunction is performed using an osteotome which is impacted with a hammer until its tip crosses through the pterygoid plates. To determine whether the disjunction has occurred and avoid complications, surgeons still rely on their proprioception.

*Methods:* The aim of this study is to validate a vibroacoustic set-up constituted by an instrumented hammer equipped with a force sensor, aiming at detecting if the disjunction has occurred. To do so, fourteen osteotomies were performed in anatomical subject jaws. For each impact, the force signal was recorded and analyzed. A classification algorithm (Support Vector Machine method coupled with a cost matrix) was developed based on indicators extracted from the signal to determine whether the disjunction had occurred. Impacts were classified into two classes, one before and the other after the osteotome had crossed the pterygoid plates. To avoid undetected disjunctions a cost matrix was added to the algorithm. The coefficients from the matrix were chosen using by minimizing the Negative Likelihood Ratio.

*Findings:* The algorithm was able to distinguish impacts before and after the disjunction with an accuracy of 96 %. Moreover, the instrumented hammer was able to detect the disjunction with a maximum delay of two impacts.

*Interpretation:* These results pave the way for the development of a per-operative decision support system for the pterygomaxillary disjunction.

233 Words

# 1 Introduction

Osteotomies are surgical procedures used in different contexts [1,2] during which a hammer is used to impact a sharp-bladed tool, called an osteotome. Le Fort osteotomy is a maxillofacial surgery aiming to correct midface deformities [3,4] by mobilizing the maxillary and/or zygomatic bones at various levels. In addition to the esthetic correction, Le Fort osteotomy can also improve breathing and eating functions.

During Le Fort osteotomy and other maxillofacial procedures such as maxillary expansion [5], the surgeon must separate the pterygoid plates belonging to the sphenoid bone, and therefore to the skull, from the palatine bone. This surgical step is called pterygomaxillary disjunction [6,7]. Each pterygoid plate is divided into medial and lateral plates forming a V-shape as described in [8]. A hook named 'hamulus' can be found at the end of each medial plate pointing towards the oral cavity. Surgeons usually insert a finger in the patient's mouth, posterior and medial to the maxillary tuberosity, to feel the hamulus through the mucosa. Then, the osteotome is impacted with a hammer until the tip of the osteotome can be felt by the finger placed in the mouth [5,7,8]. Therefore, surgeons rely on their proprioception to decide whether to stop or to continue impact on the osteotome. This subjective decision is crucial because soft tissues are present around the pterygoid plates such as the lateral and medial pterygoid muscles, vascular structures (the internal carotid [9] and the descending palatine arteries) and palatine nerves. These soft tissues must not be altered and a compromise has to found regarding the number and the energy of the impact realized on the osteotome, which requires a lot of experience. Excessive impacts may damage soft tissues and lead to complications such as hemorrhage [10,11] and nerve damage [12,13] with an occurrence reported from 4 to 9 % [10,11,14]; while insufficient impacts shall not allow to cut bone tissue. Despite these risks, osteotomies are performed without visual control of the operated zone, as they are sometimes difficult to access for anatomical or aesthetic reasons. Surgeons rely on their proprioception to perform the surgery, and their decisions are therefore driven by the sound of the hammer impact on the osteotome and/or haptic feedback.

Several techniques have been developed to improve the safety during an osteotomy. Medical imaging techniques are widely used for surgical planning of Le Fort osteotomy [3]. Pre-operative images can also be used to design patient-specific cutting guides [15]. Computer-assisted navigation systems can also be used to monitor the osteotome position and thus display the expected cutting path to the surgeon in real time [16]. Robot [17] or mixed reality [18] have been employed to position the jaw. Virtual and augmented realities have been mentioned to plan the surgery [19] and to project the cutting path on the bone [20], respectively. All the aforementioned tools are useful but have some drawbacks since they increase the operating time (which may increase infection risk [21]) and they require additional cost and space in the operating room. Moreover, these techniques are based on imaging methods which are irradiating. Piezoelectric surgery has been widely adopted for Le Fort osteotomy [22], as it enables precise cuts and avoids the risk of affecting soft tissues. However, piezoelectric surgery cannot be used for pterygomaxillary disjunction [23,24] because of the difficulty of access. So far, no tool is available to assess quantitatively whether the disjunction is complete. The development of a method which does not change the clinical routine and is capable of detecting the disjunction would be useful.

Our group has developed an impact method based on an instrumented hammer equipped with a force sensor on its impacting face to assess the acetabular cup implant stability, which was validated in animal [25], anatomical [26] and numerical [27] models. The same approach was then applied to the femoral stem *in silico* [28], in artificial bone [29], in an animal model [30] and in an anatomical model [31]. This same impact-based method was later applied to osteotomy. The force signal recorded was analyzed to monitor the osteotome dynamic response during the impact, which induces a movement of the osteotome between the hammer and the osteotomized material. A first study [32] showed that the instrumented hammer can be used to assess the sample material and thickness with an error lower than 10 %. Then, the method was applied to rhinoplasty to detect crack propagation and the transition from the nasal bone to the frontal bone *ex vivo* in animal [33] and anatomical [34] models. The same approach was applied in the context of Summers osteotomy [35] and of high tibial osteotomy in an animal model [36] and then in human bone [37]. Interestingly, the impact-based method was used to predict and detect the rupture of i) parallelepiped bovine trabecular bone samples held in a vise [38] and ii) pterygoid plates in an animal model [39]. These studies demonstrate that an instrumented hammer may be used to detect bone rupture during osteotomies. This simple method, which requires minimal equipment, could provide a quick and easy solution without requiring any changes to the surgical protocol. Although testing the method in animal tissue is a requirement before working in human tissue, such approach alone remains limited because of bone composition and geometry, which is different compared to the clinical situation.

The aim of this study is to validate the impact-based method in the case of pterygomaxillary disjunction in an anatomical model. To do so, fourteen pterygomaxillary disjunctions were performed for seven human jaws of anatomical subjects with the instrumented hammer. The osteotome position compared to the pterygoid plate was monitored using a video camera and proprioception to determine whether the osteotome tip had crossed the plate. A dedicated signal processing technique was developed and a classification algorithm aiming to detect the impact corresponding to the osteotome tip crossing the plate (which corresponds to the pterygomaxillary disjunction) was implemented and compared to the results obtained with the video camera.

# 2 Methods

## 2.1 Samples

Seven jaws were extracted by an experienced surgeon at the 'Fer-à-Moulin' surgical school, from anatomical subjects with an average age of 83.0 ± 9.5 years using a level three Le Fort osteotomy without performing the pterygomaxillary disjunction. To do so, the following steps were performed: first a coronal approach was done to expose the bone, then a frontozygomatic osteotomy, an upper zygomatic arch osteotomy, a frontonasal osteotomy and an under palpebral osteotomy were performed. The upper jaw was then extracted without performing the pterygomaxillary disjunction. Similarly as the previous study with animal model [39], soft tissues around the junction were removed until the pterygoid plates were visible. The pterygoid plate integrity was visually checked by the surgeon. The upper jaw was then placed into a square mold and a polyurethane resin (Smooth-Cast 300, Smooth-on, Pennsylvania, USA) was poured over the zygomatic arches. To obtain an optimal fixation, the resin was poured at the highest possible level without touching the plates. The resin was then fixed to the table thanks to three clamps (see Fig.1.A). This protocol was approved by the Fer-à-Moulin surgical school (Paris, France).

## 2.2 Experimental procedures

The osteotome used was similar to the instruments required for classic Le Fort osteotomy, with a 6 mm curved edge (CP304B-6, MicroFrance, Saint-Aubin-le-Monial, France). To closely replicate the clinical situation, the hammer and the osteotome were held manually.

The osteotome was slid against the maxillary tuberosity so that its curved edge blade was able to reach the pterygoid plates by getting around the maxillary tuberosity. First, the osteotome was placed at the extremity of the pterygoid plate in the *x-axis*, which corresponds to the position *#A* and *#B* in Fig.1.B, so that the entire blade rested on the bone. Note that the *x-axis* was determined based on the sample anatomy. The instrumented hammer was then used to impact the osteotome until it has crossed the pterygoid plates. The osteotome exit was monitored by a video camera placed on the other side of the plates (see Fig.1.A). This video camera had a resolution of 1920 x 1080 pixels and the image sensor format referring to the image sensor shape and size was 1/2.7". It corresponds to a sensor area of about 21.7 mm2. The camera frame rate was 30 fps (Frames per second) and the imaging dynamic range was 75 db. Additional impacts were given after the osteotome exit, to make sure that the osteotome tip was in contact with air at the end of the procedure. If the plate was not completely removed, the osteotome was translated by 6 mm (*i.e.* the length of the osteotome blade) along the x-axis, which corresponds to the position *#B* and *#C* in Fig. 1.B. The same protocol (impaction and translation of the osteotome along the *x-axis*) was repeated until the plate was completely detached from the rest of the jaw. For each osteotomy *#j*, the position taken by the osteotome will be referred to as 'osteotomy location' in what follows. The disjunction of each sample is thus composed of one or more osteotomy location, the number of locations depending on the anatomy of the specimen. The impacts were numbered starting by the second impact because before the first one the osteotome tip may not be completely in contact with the bone. The total number of impacts of the osteotomy *#j* will be noted *N(j)*.

## 2.3 Signal acquisition and processing

A piezoelectric force sensor (208C04, PCB Piezotronics, Depew, NY, USA) was screwed into the impacting face of a classical surgical hammer (32-6906-26, Zepf, Tuttlingen, Germany) weighing 260 g. This sensor had a measurement range up to 4.45 kN in compression and it was calibrated by the manufacturer. The sampling frequency and the resolution of the data acquisition module (NI 9234, National Instruments, Austin, TX, USA) used was 51.2 kHz and 24 bits, respectively. The signal corresponding to the variation of the force as a function of time was recorded for a duration of 20 ms. A part of this signal *s(t)* corresponding to a given impact of the instrumented hammer on the osteotome is shown in Fig.2.

Similarly, as in previous studies from our group [32,33,39], the signal was approximated by a Gaussian fit. For each peak *#p*, the value of the maximum peak amplitude $a_p$ was determined. The indicator $\tau$ was thus defined as the difference between the time of the maximum of the second and of the first peak, as shown in Fig.2.

The integral of each peak *#p* on the signal *s(t)*, named $L_p$, was also calculated based on $a_p$ and the Gaussian root mean square width $w_p$ with $L_p = (2\pi)^{0,5} \times w_p \times a_p$. The indicator $w_p$ is proportional to the width at -6 dB of the Gaussian function. The impulse ratio $\lambda$ was thus defined as the ratio between the first two peaks integrals, *i.e.* $\lambda = L_2 / L_1$.

For each impact *#i* ϵ [2, *N(j)*] of each osteotomy *#j* ϵ [1, 14], two other parameters, $\Delta\tau$ and $\Delta\lambda$, were calculated following: $\Delta\tau(i) = \tau(i) - \tau(i-1)$ and $\Delta\lambda = \lambda(i) - \lambda(i-1)$.

## 2.4 Data Processing

### *2.4.1 Database Creation*

Once the data was gathered and processed as mentioned previously, a class was assigned to each impact depending on the material surrounding the tip of the osteotome, similarly as what was done in [39].

- The impact was considered to belong to the class referred to as 'Bone' if the osteotome did not cross the bone plates.
- The impact was considered to belong to the class referred to as 'Crossing' when the osteotome tip was visible on the video camera monitoring the exit of the osteotome and/or detected by haptic control. This class also included the additional impacts given after the exit of the osteotome (see section 2.2).

To ensure that the correct classification was attributed to each data, we chose to optimize the detection of the osteotome exit using different approaches. First, the video camera location was carefully chosen to optimize the point of view regarding the osteotome exit. Second, an operator was monitoring the exit point from above and could ensure whether the exit of the osteotome occurred or not by placing a finger against the bone and feeling the presence of the osteotome, similarly as what is done in the clinic. Third, an experienced surgeon performed the osteotomy and used his proprioception to determine any change of bone properties, similarly as in clinical practice (*e.g.* when the osteotome is moved forward into the bone). Finally, the video was reviewed after the experiment, and a picture was taken after each impact. Comparing each image with the image taken after the exit occurred allows us to locate the exit area and thus determine any local change related to the osteotome exit. All these combined approaches allowed to optimize the determination of the impact corresponding to the exit as precisely as possible.

### *2.4.2 Statistical Analysis: a support vector machines algorithm*

Two ANOVA analyses were performed to determine whether $\tau$ and $\lambda$ were statistically different between both classes. Moreover, to combine information from $\tau$, $\lambda$, $\Delta\tau$ and $\Delta\lambda$, two variables defined by linear combinations of these four indicators were calculated following [40]. These variables, called $c_1$ and $c_2$, enable to maximize the distance between both classes. The database was thus composed of three quantities: $c_1$, $c_2$ and the class.

As in [39], a Support Vector Machines (SVM) algorithm was developed to classify the impacts and to detect the impacts labeled a 'Crossing' corresponding to impacts where the disjunction occurred. Methods based on Support Vector Machines are supervised machine learning techniques developed in the early 1990s [41] for classification and regression purposes. This technique seeks a boundary, an hyperplane, which maximizes the distance between this hyperplane and the nearest point of each class. They are currently applied in many fields and in particular, in medical engineering [42].

A soft margin SVM classifier with a linear kernel was applied as our data were not linearly separable [43]. A *k-fold* cross validation [44] was also applied in order to avoid overfitting (with k=5), leading to an algorithm which was trained on a randomly selected part of the database and tested on the other unknown part of the database. This procedure was performed five times and the accuracy reported below corresponds to the average of the accuracy values measured for each loop.

The classification algorithm assigned a 'predicted class' to each impact of the database, which can be compared with the class attributed using the video camera, referred to as 'true class'. Since we aimed at detecting the osteotome crossing, the positive (respectively negative) event was defined as an impact belonging to 'Crossing' (respectively 'Bone') class. Therefore, a True Positive (TP) (respectively a True Negative (TN)) event corresponded to an impact where the predicted and actual classes were both 'Crossing' (respectively 'Bone'). Conversely, a False Positive (FP) (respectively a False Negative (FN)) event was defined as an impact where the predicted class was 'Crossing' (respectively 'Bone') whereas the true class was 'Bone' (respectively 'Crossing').

### 2.4.3 Optimization of the classification algorithm

FN events should be avoided as they represent undetected osteotome crossing. To reduce FN events, a cost matrix was added into the SVM algorithm described above. This cost matrix, $\sigma$, was defined as in [45]:

$$\sigma = \begin{bmatrix} \sigma_{BB} & \sigma_{CB} \\ \sigma_{BC} & \sigma_{CC} \end{bmatrix} \quad (1)$$

The subscript B (respectively C) refers to the class 'Bone' (respectively 'Crossing') and each coefficient $\sigma_{lm}$ is the relative weight applied for an impact with a true class '*l*' and a predicted class '$m$'. The cost matrix coefficients were optimized using the same method as [39,46] based on a ratio called Negative Likelihood Ratio (NLR), defined in [47] and given by:

$$NLR = \frac{FN/(TP + FN)}{TN/(FP + TN)} \ (2)$$

The optimization aimed at determining coefficients reducing the value of the NLR, leading to a decrease of the FN events, while preserving a maximum number of TN. Several initial coefficients $\sigma_{CB}$ and $\sigma_{BC}$ were tested and a Nelder–Mead method [48] was used to determine the coefficient optimal values allowing to obtain a local minimum of the NLR ratio.

To evaluate the classification algorithm performance, some classical metrics such as sensitivity, specificity, ROC curve and AUC were calculated.

All data were processed with Matlab (MATLAB, R2017b, MathWorks, Natick, Massachusetts, USA).

# 3 Results

## 3.1 Statistical analysis of the database

For all samples, an average number of 1.41 ± 0.52 osteotomy locations were carried out in order to completely detach the plates from the rest of the jaw, *i.e.* to achieve a complete disjunction. A complete disjunction consists on average of 39.79 ± 19.41 impacts. Table 1 displays the average impact number for each location as well as each location occurrence. The impact number tends to decrease between the first and the second location because the pterygoid plates are already weakened by the first osteotomy.

## 3.2 Result for a representative sample

Figure 3 shows the variation of $\tau$ and $\lambda$ as a function of the impact number during the osteotomy procedure of the sample #8D. Stars indicate impacts classified by the prediction algorithm as 'Crossing' and circles correspond to incorrect prediction of the algorithm.

The osteotomy of sample #8D is composed of only one location. Figure 3.A shows the behavior of $\tau$, which is qualitatively similar to the one obtained in the animal study [39]. Before the crossing of the osteotome tip, the values of $\tau$ are relatively constant around a value of 0.955 ± 0.106 ms and then suddenly increase when the osteotome crosses the plates (impacts #60 to #63) until it reaches a maximum value of 3.185 ms. Figure 3.B shows that despite a sharp decrease of $\lambda$ for the impact corresponding to the crossing of the osteotome (see impacts #60 to #64), the general behavior of $\lambda$ is more scattered. $\lambda$ is not stable when the osteotome has not crossed the plates and a sharp decrease can be observed between impacts #35 to #38 and on the impacts #14, #48 and #51. For sample #8D, the classification algorithm only leads to one FP event (see circle on impact #59) and to no FN event.

### 3.3 Classification results

Two ANOVA analyses performed on all the data confirmed that $\tau$ and $\lambda$ are significantly different before and after the plate disjunction as determined by the video camera assessment. The values of $\tau$ (respectively $\lambda$) are significantly higher (respectively lower) for impacts belonging to the 'Crossing' class compared to the ones belonging to the 'Bone' class with $p < 0.001$ and $F = 125.3$ (respectively $p < 0.001$ and $F = 294.3$).

The cost matrix used after the application of the optimization method described in section 2.4.3 reads:

$$\sigma = \begin{bmatrix} 0 & 3.29 \\ 2.10 & 0 \end{bmatrix} \qquad (3)$$

The classification algorithm performances with and without the use of the cost matrix are detailed in Tab.2.

The ability to correctly identify TP (respectively TN) events, known as the sensitivity (respectively specificity), for this prediction algorithm with the cost matrix is equal to 84 % (respectively 99 %). The specificity is higher than the sensitivity, which is consistent with the application of the cost matrix penalizing FN events. The AUC of the ROC curve (see Fig.4) is 0.971, indicating that the algorithm is efficient. The number of TP (respectively TN) events for the classification algorithm is 48 (respectively 493).

Figure 5 represents all 557 data corresponding to all impacts displayed in the canonical space. The green crosses represent impacts belonging to the 'Crossing' class and the red dots indicate impacts belonging to the 'Bone' class. The 'Crossing' (respectively 'Bone') prediction spaces is displayed in light gray (respectively dark gray). The boundary between the two spaces with (respectively without) the cost matrix is indicated by a black (respectively white) line.

## 4 Discussion

The novelty of the present study is to prove that the pterygomaxillary disjunction performed in a cadaveric model can be detected thanks to an instrumented hammer used in combination with a classification algorithm (SVM method with a cost matrix). This detection can be useful in many maxillofacial surgeries because it can avoid unnecessary impacts. The main difference with the previous animal study lies in particular in bone tissue composition and geometry.

### 4.1 Physical interpretation of the results

The force signal shown in Fig.2 is qualitatively consistent with the signals predicted by a simple analytical model developed in [32]. Several peaks can be observed, the first one corresponding to the initial impact between the hammer and the osteotome. The following peaks may be explained by the osteotome rebounds between the hammer and the osteotomized material.

The indicator $\tau$ was linked to the material rigidity surrounding the osteotome tip in a previous study [32]. Figure 3.A shows that an increase of $\tau$ was obtained when the osteotome crossed the bone plate. This result is coherent with previous studies with bone plates [38] and with the animal model for pterygomaxillary disjunction [39]. The decrease of $\tau$ when the osteotome crosses the bone plate may be explained by a loss of rigidity of the bone-osteotome system.

The other indicator $\lambda$ has been linked to the viscoelasticity of the material surrounding the osteotome tip in [32]. Although $\lambda$ tends to decrease when the osteotome crosses the plates, some other unexpected variations can be observed (*e.g.* impacts #35 to #38 in Fig.3.B). The relationship between $\lambda$ and the bone plate rupture has not been investigated here and will be left for future studies.

## 4.2 Anatomical subject characteristics

The previous study [39] demonstrated the ability of the instrumented hammer to detect the disjunction in animal tissue. While the demonstration in animal tissue before working with human tissue is an ethical requirement, the animal bone mechanical properties are different compared to human bone. The introduction of human models changes various aspects.

Lamb bone and human pterygoid plates are different in terms of geometry and of biomechanical properties. The lamb pterygoid plates consist of a single flat plate (*i.e.* one on each side of the head), parallel to each other with a pit filled with soft tissue in the middle. They also have an oval cavity filled with soft tissue (see Fig.1 in [39]). The human plates are composed of two plates forming a V-shape on each side. There is a lot of soft tissues around them but no oval cavity. Some other geometrical differences can be observed such as the hook mentioned in the introduction.

Other geometric characteristics of the pterygoid plates could be quantified. The plate width (distance between the edge and the end of a plate in the x-axis indicated in Fig.1.B) is different between the human and the lamb model. In the previous study [39], this distance was always larger than 12 mm whereas it was around 9 mm on the human subjects of the present study. These results explain why the average number of locations per plate is smaller herein compared to the previous study (1.41 vs 3.42 locations per osteotomy with the anatomical subjects and lamb studies, respectively).

Moreover, although the plate thickness couldn't be assessed in the anatomical subject study, it has been reported to be around 7.7 mm according to [49]. In the previous study with lamb, the plate thickness fluctuated along the plate and was always under 4.8 mm. These difference in the plate thickness could explain why the average number of impacts given by location (*e.g.* 30.64 ± 14.3 for the first location) in the human model is higher compared to the lamb model (*e.g.* 14.60 ± 12.82 for the first location).

Although the use of anatomical subjects allows to test the method in a situation closer to the clinics compared to an animal study [39], the age of the anatomical subjects (83.0 ± 9.5 years) remains a limitation of this study because it is higher compared to the average age of the patients needing Le Fort osteotomy (from 19.5 up to 28.5 years as reported in [50]). Using elderly subjects is common for maxillofacial studies with anatomical subjects [51], but it could have an impact on the signal characteristics and thus on the classification. In addition to the reduced bone quality, the age of the subjects also implies that they are partially or completely edentulous, which could have an impact on the pterygoid plates bone quality. To reduce the difference between the model and the clinical situation, the donors who were completely edentulous and who did not wear any dental prosthesis were excluded. The surgeon was also able to give a subjective evaluation based on his own proprioception and experience. No quantitative bone quality measurement could be performed in the surgical school and we were not allowed to move the samples to do additional tests elsewhere. This is a limitation of the present study and future works should include a method to evaluate bone properties.

Finally, the freeze-thaw cycles of the subject have a minimal influence on the bone properties as mentioned by [52,53]. The freeze-thaw cycles effect on the signal and the indicators presented herein was not checked in this study. This is a current limitation inherent to the work on anatomical subjects, and it should be checked in future studies

## 4.3 Toward more accurate models

As described in the previous section, some limitations of the study are inherent to the use of anatomical subjects. Some choices regarding the experimental set-up induce a significant difference with the clinical situation. These different simplifications were considered in order to be able to quantify the performances of the method to detect the disjunction in a controlled environment.

First, we chose to remove the jaw from its anatomic position in order i) to access the osteotome exit area, which is otherwise very difficult to reach and ii) to observe and feel the osteotome exit and iii) to check the integrity of the plates.

Second, soft tissues were removed to i) check if the plate had a defect coming from a previous surgery or from another event of the donor life and ii) derive a better detection of the osteotome exit from the camera recording or by proprioception. The influence of soft tissues was found to be negligible in a previous study [54] on total hip arthroplasty. However, further studies are needed to determine their influence in the case of osteotomies as the osteotome tip will be in direct contact with the soft tissues at the end of the osteotomy which was not the case in the total hip arthroplasty.

Future studies must be carried out in a situation closer to the one of the operating room and model becoming closer to the clinical situation step by step, including, but not limited to, a study on an extracted jaw with soft tissues, and then a jaw in its correct anatomical position with the soft tissues and then a clinical trial. For all these studies the osteotome exit and thus the disjunction will need to be checked using an independent method such as an imaging method.

To use this method in clinical practice there are still some technical considerations. First, the use of an acquisition system and a computer would make the operating room routine more complicated, so the final design of the hammer should include the sensor and the entire processing device within the hammer. The surgeon would then be alerted to an event by a light and/or sound signal. Second, the hammer must be able to be sterilized using conventional techniques for surgical instruments. These are future steps for potential surgical translation and are left for future studies.

### 4.4 Other limitations

Other limitations should be considered. First, some standardization rules were applied herein in order to optimize the reproducibility of the measurements. The operator was asked to impact the osteotome with a force comprised between 200 and 2000 N. This force range allows the progression of the osteotome, while ensuring that it does not cross too fast through the bone. Please note that impacts were given with a mean force of 1454.3 ± 1381.6 N. Regarding the osteotome placement, the only instruction was to place it against the maxillary tuberosity similarly to clinical practice. Please note that although the force value of the impacts was recorded as indicated above, it was not used in the classification algorithm as it is a parameter dependent on the operator striking the osteotome.

Only one operator (an experienced maxilla-facial surgeon) impacted the osteotome in this study. However, we would like to highlight that this study is the first to demonstrate the ability of the instrumented hammer to detect the pterygomaxillary disjunction in human tissue. Moreover, in other surgical fields such as total hip arthroplasty [29] or rhinoplasty [33,34], the instrumented hammer was used by various operators and still gave consistent results. This point is a limitation of the current study and is therefore left for future studies.

Second, k-fold cross-validation is a known method to avoid overfitting since the prediction algorithm is trained on a randomly selected subset of data (four-fifths of the dataset) and tested on the rest of the dataset, which is unknown [55]. Although the use of k-fold cross-validation, there is still a risk that the limited number of impacts could result in overfitting. This is a current limitation of the study, and a validation study on an independent dataset would be interesting to confirm the results Although data augmentation is a valid method, we would rather work on experimental data and thus another campaign in another center would be the next step for the development of the instrumented hammer for the pterygomaxillary disjunction.

Finally, we chose to use a classification algorithm (SVM method) to distinguish the impact occurring before (*i.e.* impacts labeled as 'Bone') and after (*i.e.* impacts labeled as 'Crossing') the disjunction. Although the cost matrix improves the algorithm accuracy, some misclassification remains. As shown in Tab. 2, the classification algorithm is associated with 7 FP (impacts belonging to the class 'Bone' and classified as 'Crossing') and 9 FN (impacts belonging to the class 'Crossing' and classified as 'Bone'). For 8 out of 9 FN event, the crossing of the osteotome was correctly detected at the following impact. For 1 FN event, the crossing of the osteotome was correctly detected two impacts later.

Regarding the 7 FP events, for 2 of them, the osteotome crossed the bone one impact after the FP event. For another FP event, the osteotome crossed the bone two impacts after the FP event. For 3 FP events, the osteotome had partially crossed the bone on the FP event. All these FP events are less problematic than FN events because they constitute warnings before the crossing of the osteotome. For one last remaining FP, $\tau$ unexpectedly increases and $\lambda$ unexpectedly decreases 5 impacts before the real crossing of the osteotome. A micro-crack or another defect could have occurred in this sample.

## 5 Conclusions

This study focuses on the use of an instrumented hammer to detect pterygomaxillary disjunction (pterygoid plate separation from the rest of the skull). Detecting the impact where the osteotome crosses trough the bone plate is useful for many maxillofacial surgery (*e.g.* Le Fort osteotomy) as it can avoid unnecessary impacts which can lead to complications such as hemorrhages. To do so, osteotomies were performed on upper jaw extracted from anatomical subjects. For each impact, the force variation as a function of time was recorded and analyzed using a dedicated procedure. Four indicators were extracted from the signal and were used to distinguish impacts given before and after the crossing of the osteotome, which corresponds to the disjunction. To do so, a SVM classifier with a cost matrix was employed and led to correctly detect the crossing of the osteotome with an accuracy of 96 % and a maximum delay of two impacts. These results pave the way for the development of an intraoperative decision support system for maxillofacial surgeons.

Acknowledgments: The authors would like to thank Augustin Daridon, Carla Bédart and Aglaé Marcandella for their help in obtaining experimental results.

Author Contributions: Conceptualization, M.B.d.N. and G.H.; methodology, G.H. and G.R.; software, M.B.d.N.; validation, M.B.d.N.; formal analysis, M.B.d.N. and G.T.; investigation, M.B.d.N.; resources, G.H.; data curation: M.B.d.N. and G.T.; writing—original draft preparation, M.B.d.N.; writing—review and editing, G.H., C.H.F.L., G.T. and G.R.; visualization, M.B.d.N.; supervision, G.H.; project administration, G.H.; funding acquisition, G.H. All authors have read and agreed to the published version of the manuscript.

Funding: This project has received funding from the project OrthAncil (ANR-21-CE19-0035-03), from the project OrthoMat (ANR-21-CE17-0004), from the project MaxillOsteo (ANR-ANR-24-CE17-2236), from the project MoDyBe (ANR-ANR-24-CE17-2236). This project has received funding from the European Research Council (ERC) under Horizon 2020 (grant agreement # 101062467, project ERC Proof of Concept Impactor).

Conflict of interest: The authors declare that they have no conflict of interest.

## Figure Legends

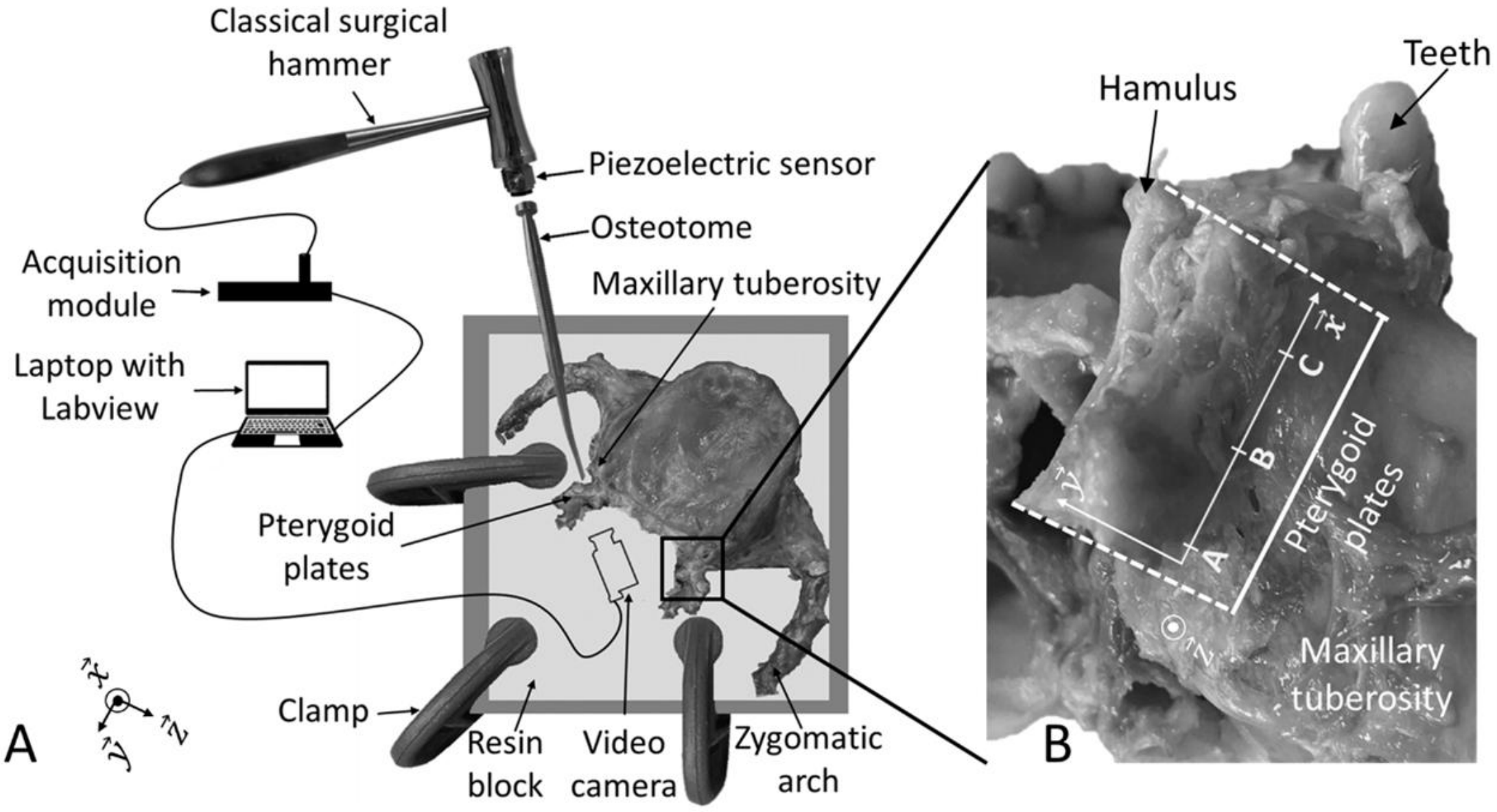

Fig 1: Experimental configuration during the osteotomy on the anatomical subjects. A: Experimental setup showing the position of: the camera, the sample, the osteotome and the instrumented hammer. The impacts realized with the hammer were performed in a direction parallel to the axis of the osteotome; B: Zoom on the pterygoid plates each letter, separated by 3 mm, corresponds to the location of the osteotome considered during the experimental protocol. The osteotome blade was slid along the maxillary tuberosity and placed perpendicular to the *xy* plane.

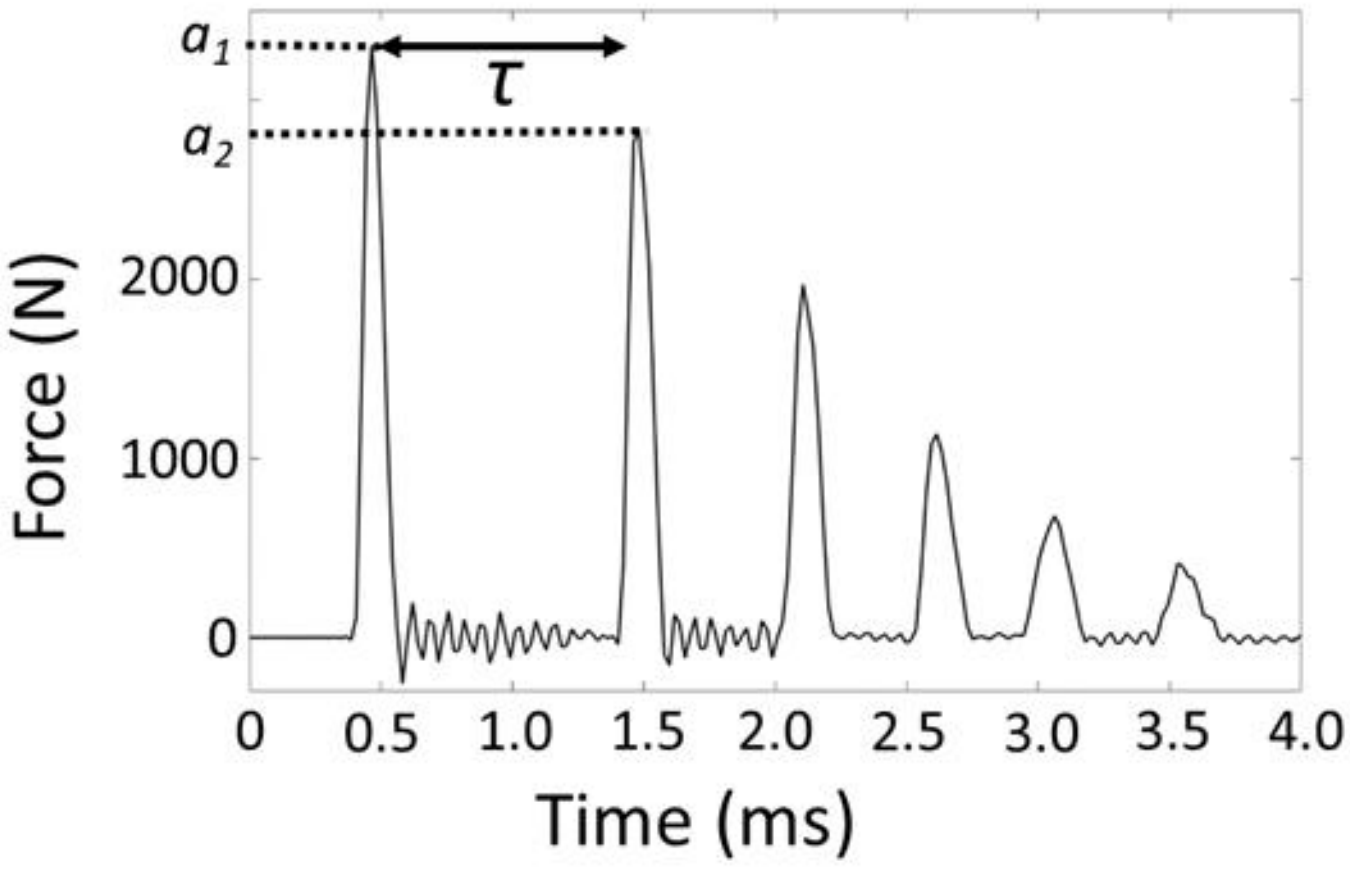

Fig 2: Part of the signal *s(t)*, variation of the force as a function of time, corresponding to a given impact of the instrumented hammer on the osteotome. The indicator $\tau$ denotes the time difference between the two first peaks of the signal. $a_1$ and $a_2$ correspond to the values of the maximum amplitude of the first and second peak, respectively.

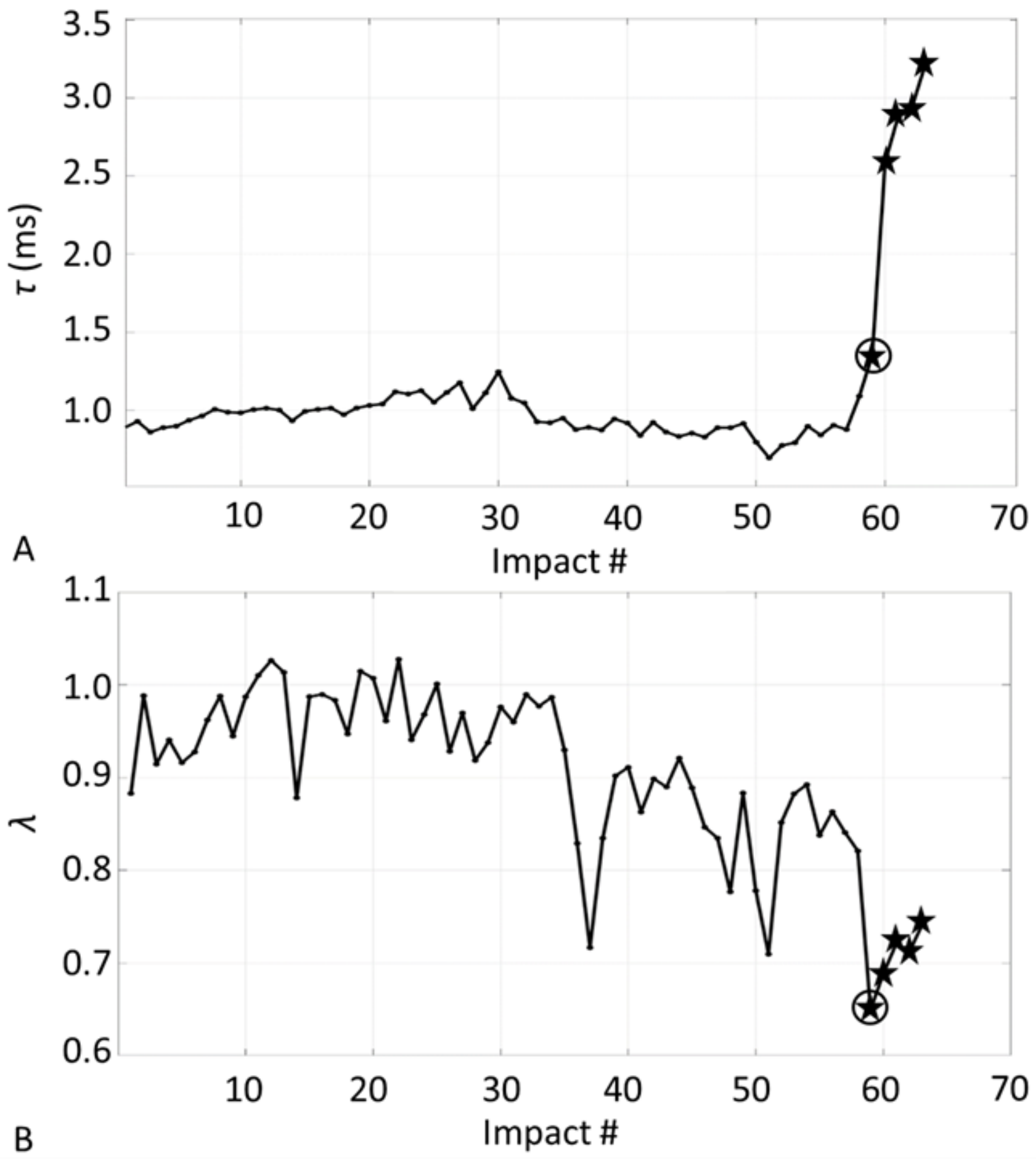

Fig 3: Result obtained for the sample #8D. A: Variation of $\tau$ as a function of the impact number; B: Variation of $\lambda$ as a function of the impact number. In both figures, the stars correspond to the impact classified as 'Crossing' by the classification algorithm and the prediction error (here a False Positive (FP) event) is indicated with a circle. The other impacts thus correspond to the impact classified as 'Bone'.

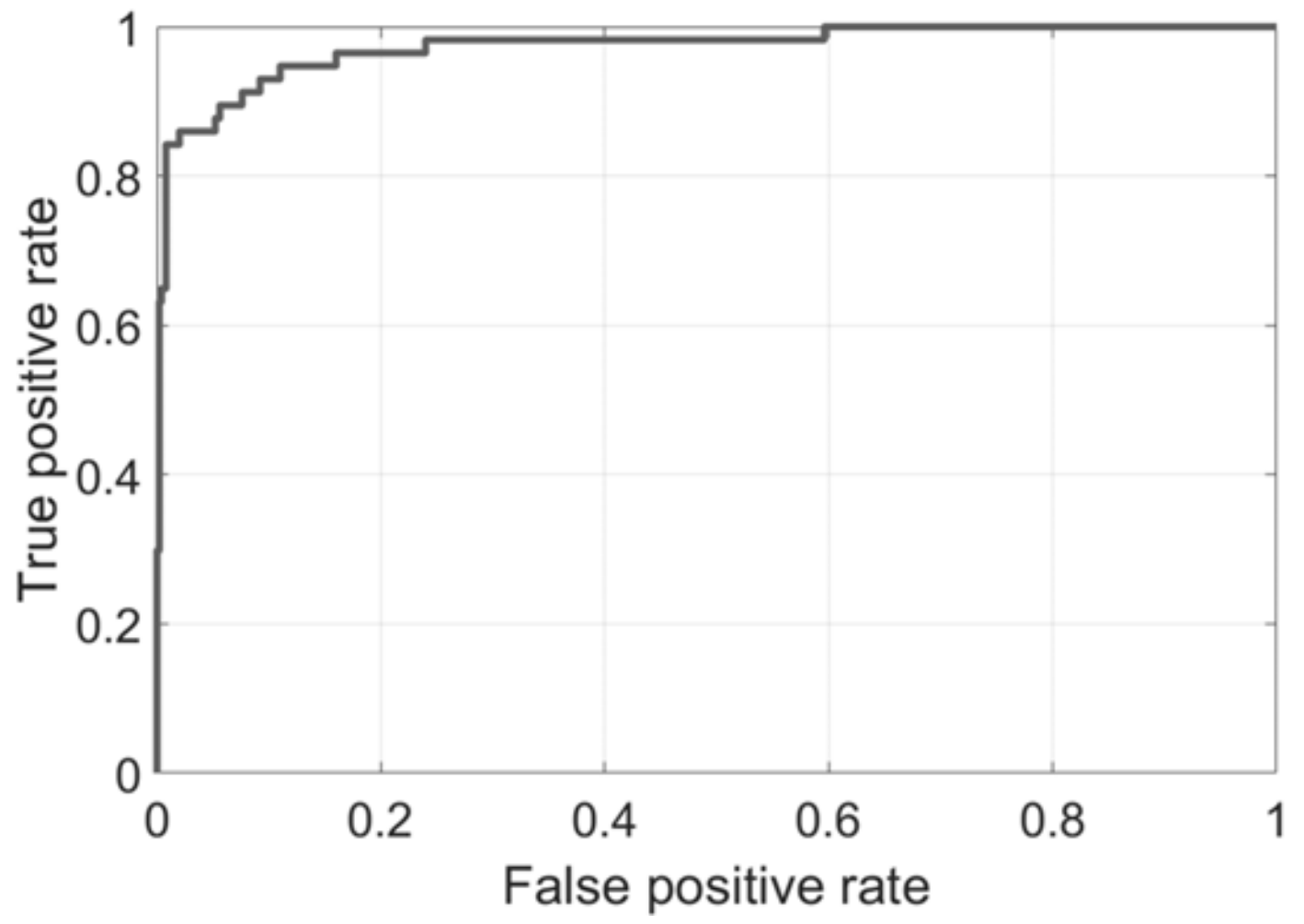

Fig 4: ROC curve for the prediction algorithm with the cost matrix.

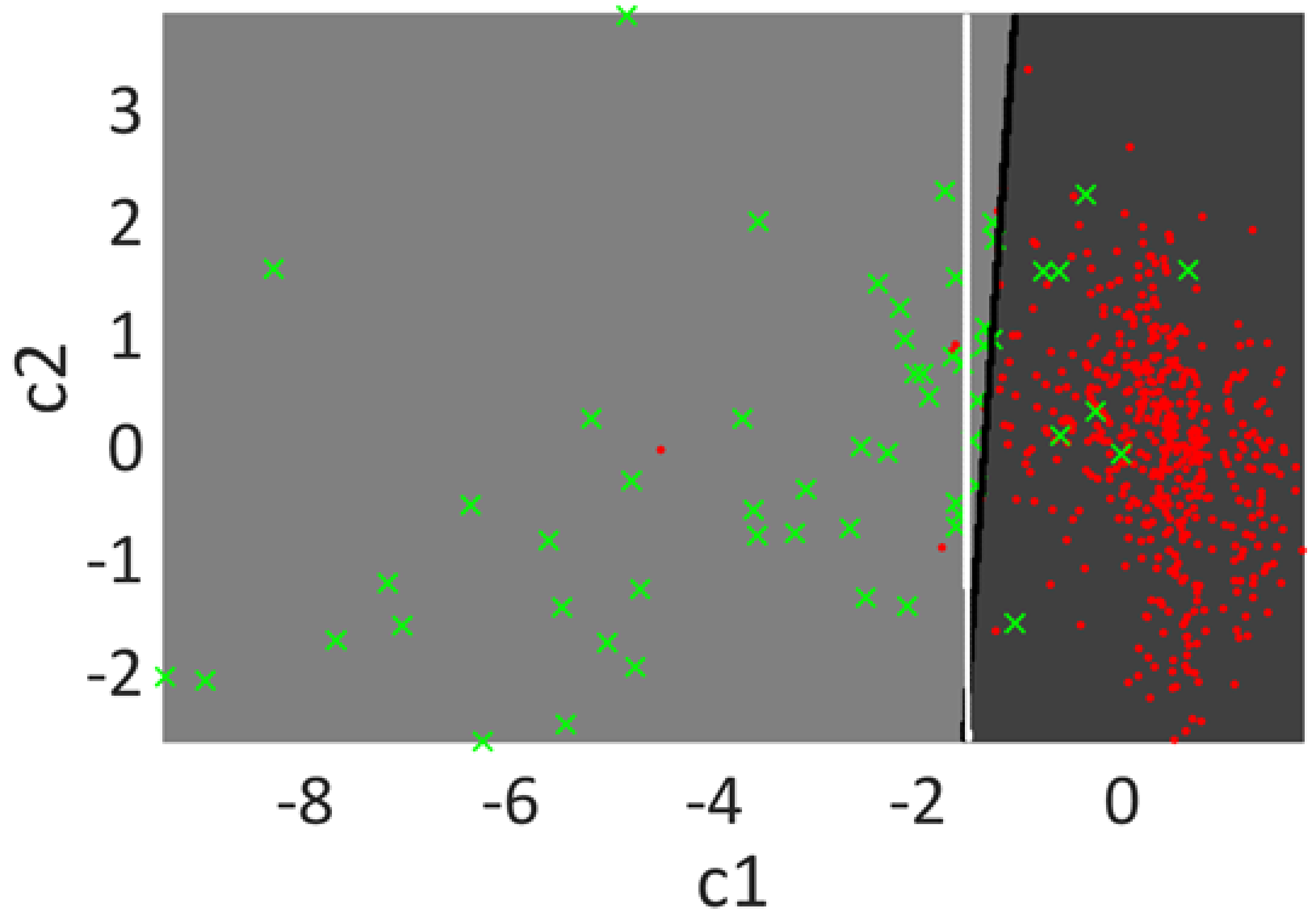

Fig 5: Classification algorithm results in the canonical variable space. The red dots and green crosses represent the impacts belonging to the ‘Bone’ and ‘Crossing’ classes, respectively. The two different spaces represent the result of the classification algorithm. The light gray (respectively dark gray) space corresponds to the ‘Crossing’ (respectively ‘Bone’) space prediction. The black (respectively white) line corresponds to the boundary between the two spaces with (respectively without) the cost matrix.